
%
%
%
\input phyzzx
\unnumberedchapters


%

%
%

%

%
\def\refoutlw{\par\penalty-400\vskip\chapterskip
   \spacecheck\referenceminspace
   \ifreferenceopen \Closeout\referencewrite \referenceopenfalse \fi
   \line{\bf\hfil References\hfil}\vskip\headskip
   \input \jobname.refs
   }
%
%
\def\a{\alpha}
\def\b{\beta}
\def\d{\delta}
\def\ga{\gamma}
\def\m{\mu}
\def\n{\nu}
\def\r{\rho}
\def\s{\sigma}
\def\ee{\epsilon}
\def\gm{\Gamma}
\def\La{\Lambda}
%
%

\def\VV{\scriptscriptstyle V}

\def\SCS{S_{\rm CS}}
\def\SGF{S_{\rm GF}}
\def\cv{c_{\VV}}

\def\idx{\int \! d^3\!x\>}

%

\mathsurround=2pt
\rightline{FTUAM 94/8}
\rightline{NIKHEF-H 94/14}
\rightline{UPRF 94/395}
\date{}

%
\REF\Witten{E. Witten, Commun. Math. Phys. {\bf 121} (1989) 351.}
\REF\shift{L. Alvarez-Gaum\'e, J.M.F. Labastida and A.V.
          Ramallo, Nucl. Phys. {\bf 334B} (1990) 103.
\nextline M. Asorey and F. Falceto, Phys. Lett. {\bf 241B} (1990) 31.
\nextline C. P. Martin, Phys. Lett. {\bf 241B} (1990) 513.
\nextline G. P. Korchemsky, Mod. Phys. Lett. {\bf 6A} (1990) 727.
\nextline D. Birmingham, R. Kantwoski and M. Rakowski, Phys. Lett.
          {\bf 251B} (1990) 367.
\nextline G. Leibbrandt and C. P. Martin, Nucl. Phys. {\bf 377B} (1992)
          593.}
\REF\Freed{D.S. Freed and R.E. Gompf, Commun. Math. Phys. {\bf 141}
          (1991) 79.}
\REF\universality{G. Giavarini, C. P. Martin  and F. Ruiz Ruiz, Phys.
Lett. {\bf 314B} (1993) 328.}
\REF\Asorey{M. Asorey, F. Falceto, J. L. L\'opez and G. Luz\'on,
{\it Universality and Ultraviolet Regularizations of Chern-Simons
Theory}, DFTUZ 93.10 preprint.}
\REF\nonlocality{D. A. Eliezer and R. P. Woodard, Nucl. Phys.
{\bf 325B} (1989) 389.}
\REF\GMR{G. Giavarini, C.P. Martin and F. Ruiz Ruiz, Nucl. Phys.
{\bf 381B} (1992) 222.}
\REF\Collins{J.C. Collins, {\it Renormalization} (Cambridge
University Press, Cambridge, 1984).}
\REF\meaning{G. Giavarini, C. P. Martin  and F. Ruiz Ruiz, Phys. Rev.
{\bf 47D} (1993) 5536.}
%

\titlepage

\title{{\seventeenbf Shift versus no-shift in local regularizations of
Chern-Simons theory}}

\author{ G. Giavarini }
\address{{\it Libera Universit\`a della Bassa Ovest, Villa Baroni 3,
              43107 San Secondo P.se, Italy \break
              and \break
              Dipartimento di Fisica dell'Universit\`a di Parma,
              Viale delle Scienze, I-43100 Parma, Italy}}

\author{ C. P. Martin}
\address{{\it Departamento de F\'\i sica Te\'orica,  C-XI,
              Universidad Aut\'onoma de Madrid \break
              Cantoblanco, Madrid 28049, Spain}}

\author{ F. Ruiz Ruiz}
\address{{\it NIKHEF-H, Postbus 41882, 1009 DB Amsterdam,
              The Netherlands}}

\vskip 2 true cm

\noindent
We consider a family of local BRS-invariant higher covariant
derivative regularizations of $SU(N)$ Chern-Simons theory that do not
shift the value of the Chern-Simons parameter $k$ to $k+\,{\rm
sign}(k)\,\cv$ at one loop.


\endpage

\pagenumber=2

The problem of the uniqueness of the shift $k \to k+\a$ of the bare
Chern-Simons parameter $k$ in perturbative quantization of
Chern-Simons theory has been open to debate for some years. All
explicit computations of $\a$ at one loop performed as yet
[\Witten-\universality] using local gauge invariant regulators have
given the same result, namely $\a={\rm sign}(k)\,\cv.$ The uniqueness
of the result and the very different nature of the regulators employed
led to the conjecture [\universality] that gauge invariance, if
preserved at the regularized level, chooses for the shift the value
$\a={\rm sign}(k)\,\cv$ at one loop. Further evidence for this
conjecture came from the fundamental role that gauge invariance plays
in establishing the connection between Chern-Simons theory and
two-dimensional current algebra [\Witten]. In addition, if the
bare Chern-Simons parameter has the same meaning in perturbative
quantization as in canonical quantization, the shift $\a={\rm
sign}(k)\,\cv$ is a necessary condition for the semiclassical
evaluation of the partition function [\Freed] to agree with the
nonperturbative value computed using surgery techniques [\Witten].

In ref. [\universality] we introduced a local biparametric family of
gauge invariant regularizations based on higher covariant derivative
actions. The actions we considered contained parity odd as well
as parity even terms, and had a large-mometum behaviour dominated by
parity even terms. We found that the shift $\a$ was independent of the
family paramters and equal in all cases to $\a={\rm sign}(k)\,\cv.$
Very recently, the authors of ref. [\Asorey] have devised another
family of regularization methods, still based on gauge-invariant
higher covariant derivative actions, which allow for shifts different
from $\a={\rm sign}(k)\,\cv.$ More precisely, for $k$ positive, they
have obtained the three following types of radiative corrections $\a:$
\itemitem{\rm (i)} $\a =\cv,$ whenever the large-momentum leading term
of the regularized action is parity even.
\itemitem{\rm (ii)} $\alpha=s\cv,$ with $s=0,2,$ whenever the
large-momentum leading term of the regularized action is parity odd.
The value of $s$ depending on the sign of such term.
\itemitem{\rm (iii)} $\a =r\cv$, with $r$ a real number, whenever
the large-momentum leading term of the regularized action is a linear
combination of parity even and parity odd terms. The value of $r$
depending on the coefficients of this linear combination.

\noindent The result in (i) agrees with all previous computations
[\Witten-\Freed], and in particular with those in [\universality].
However, the results in (ii) and (iii) contradict the idea of a unique
shift for all gauge invariant regulators, hence we think they deserve
some consideration. They have been obtained using regularizations based
on actions that contain nonlocal interactions, and is well known that
theories whose actions contain nonlocal interactions do not generally
have the same perturbative properties as theories whose actions only
contain local interactions [\nonlocality].  One could argue that the
differences between regularizations involving nonlocal interactions
and regularizations with only local interactions vanish as the
regulator is removed. This, however, is not obvious since the
unregularized theory is not finite by power counting. We thus find it
necessary to reproduce the results in (ii) and (iii) by employing
regularizations whose Feynman rules come from local actions. This way,
any perverse effect due to the nonlocal structure of the
regularization employed in ref. [\Asorey] can be ruled out. The
purpose of this short note is to reproduce the type of one-loop
radiative corrections in (ii) by using a fully BRS invariant
regularization method whose action is local.

The classical Chern-Simons action for gauge group $SU(N)$ in the
Landau gauge $\partial\!A^a=0$ is given by
$$
S = \SCS + \SGF \>,
$$
where
$$
\eqalign{
& \SCS = -\,{ik\over 4\pi} \idx \ee^{\m\n\r}
    \left( {1\over 2}\,\, A^a_\m \partial_\n A^a_\r
    + {1 \over 3!}\, f^{abc}A^a_\m A^b_\n A^c_\r \right) \cr
& \SGF = \idx  \!\left[ -b^a\partial A^a
        + \big(J^{a\m} - \partial^\m \bar c^a\big)
          \big(D_{\m} c^a\big)
        - {1\over 2}\,f^{abc}\, H^a c^b c^c\, \right] \> .\cr}
$$
Here the notation is as in ref. [\GMR], and for convenience we
choose $k$ to be positive. To regularize the theory, we add to $S$
the following higher covariant derivative terms:
$$
S_\La = S
+ {k\over 4\pi}\idx \!\left[\, {1\over 4\La} \> F^a_{\m\n}F^{a\,\m\n}
      - {iv \over 2\La^2} \> \ee^{\m\n\r}
        F^a_{\m\s} (D_\n F_\r^{~\s})^a \,\right] \>,
\eqn\SLa
$$
with $\La$ the higher covariant derivative mass and $v$ an arbitrary
real parameter. For all $v$ different from zero, the large-momentum
leading term in $S_\La$ is parity odd, so eq. \SLa\ provides a
family of local actions satisfying the parity requirement in (ii)
above. The adding of the higher covariant derivative terms in eq.
\SLa\ does not completely regularize the theory, since there is still
a finite number of Feynman diagrams divergent by power counting. To
regularize these diagrams, we use 't$\!$Hooft-Veltman's dimensional
regularization prescription for theories involving parity violating
objects [\Collins] (see ref. [\GMR] for details). Since this
prescription manifestly preserves gauge invariance and the action
$S_\La$ is BRS invariant, we end up with a local BRS-invariant
regularization method which combines the mass $\La$ and the
dimensional regulator $D.$ Here we are concerned with radiative
corrections to the bare Chern-Simons parameter, so we will be working
in a renormalization scheme characterized by $k_{\rm renormalized} =
k_{\rm bare}=k,$ or in other words by the renormalization constant for
$k$ being simply $Z_k=1.$ We define renormalized Chern-Simons theory
in this scheme as the limit $\La \to
\infty$ of the limit $D\to 3$ of the dimensionally regularized theory
whose classical action is $S_\La$.

We recall that up to one loop the local part of the renormalized
effective action, obtained as a solution of the BRS identities, can be
cast as [\GMR] [\meaning]
$$
\gm^{\rm loc} = - \,{i(k+\a)\over 4\pi}\, \SCS
    + \idx \left\{ -b^a\partial A^a + \Delta \left[\> \b \,(J^{a\m} -
\partial^\m \bar c^a )A^a_\m - (1+\ga)\, H^a c^a \, \right] \right\}\,.
\eqn\effective
$$
Here $\a,~\b$ and $\ga$ are coefficients of order $\hbar,$ and
$\Delta$ is the Slavnov-Taylor operator for the theory [\GMR]
[\meaning]. The effective action contains two types of radiative
corrections: gauge invariant radiative corrections, labeled by $\a,$
which make the shift, and gauge dependent radiative corrections,
labeled by $\b$ and $\ga.$ The coefficients $\a,~\b$ and $\ga$ can be
uniquely determined from the vacuum polarization tensor
$\Pi^{ab}_{\m\n},$ the ghost self-energy $\Omega^{ab}$ and the
$Hcc\!$-vertex $V^{abc}.$ All we have to do then is to compute the
limit $\La\to\infty, ~D\to 3$ of these three Green functions. This is
done by following the same steps as in ref.  [\universality], the only
difference being that the $D\!$-dimensional gauge field propagator now
reads
$$
D^{ab}_{\m\n} (q)= {4\pi \over k} \> {\La^2\,\d^{ab}
                             \over  q^2\, P(q^2,\La;v)}
      ~ \left[ \,(\La^2 + v q^2)\,\ee_{\m\r\n}q^\r
             + \La\, (q^2 g_{\m\n} - q_\m q_\n) \right] \, ,
$$
where $P(q^2,\La;v)$ has the form\foot{Note that $P(q^2,\La;v)$ is
positive definite for all $v$ and all $q^\m,$ which ensures that power
counting holds.}
$$
P(q^2,\La;v) = {(\La^2 + v q^2)}^2 + \La^2\,q^2
$$
and $\ee_{\m\r\n}$ in $D$ dimensions is understood in the
't$\!$Hooft-Veltman sense [\GMR]. Afer some calculations, we obtain
$$
\Pi^{ab}_{\m\n} (p) = \cv \, J(v)\>\ee_{\m\r\n} p^\r\,\d^{ab}
\qquad
\Omega^{ab}(p) = - \,{\cv\over k}\> I(v)\>p^2 \,\d^{ab}
\qquad
V^{abc}(p_1,p_2)=0 \,,
\eqn\Green
$$
where $p^\m,\>p_1^\m$ and $p_2^\m$ are external momenta, and $J(v)$ and
$I(v)$ are given by
$$
\eqalign{
& J(v) = - \,{2\over 3\pi} \int_{-\infty}^{\infty} \! dt \>
       { 3\,v^3t^6 - 8\,v^2t^4 - (13\,v + 5)\,t^2 - 2
                                          \over P^2(t^2,1;v)} \cr
& I(v) = {2\over 3\pi} \int_{-\infty}^{\infty} \! dt \>
       { 1 \over P(t^2,1;v)} ~. \cr}
$$
Note that in the renormalization scheme we are working there are no
restrictions on the renormalization constants $Z_\phi$ for the fields
$\phi=A^a_\m, b^a\!, \bar{c}^a\!,\ldots$ To account for this
arbitrariness, we replace the fields $\phi$ in the effective action
\effective\ with $(1+z_\phi)\phi,$ where the $z_\phi{\rm 's}$ are
abitrary coefficients of order $\hbar$ satisfying the BRS identites
$z_A+z_b=0$ and $z_A+z_{\bar c}= z_c+z_H.$ The results in eq. \Green\
then imply that
$$
\a = \cv \, \left[ \>J(v) - 2\,I(v)\>\right] \qquad ~
\b + z_A = {4\over 3}\>{\cv \over k}\,I(v) \qquad ~
\ga + 2\,z_c + z_H = 0\>.
$$
Some simple algebra to combine $J(v)-2I(v)$ into a single integral and
Cauchy's residue theorem finally lead to the following result for
$\a:$
$$
\a \,=\, \left\{ \eqalign{ 0 \qquad~\> & {\rm if}~v>0 \cr
                           2\cv\qquad    & {\rm if}~v<0. \cr} \right.
\eqn\noshift
$$
We thus see that our regularization method, based on actions with only
local interactions, reproduces the one-loop shifts in (ii) above.

Here we have calculated the shift $\a$ for $k>0$. The value of $\a$
for $k<0$ can be retrieved from our previous computations in the
following way. Assume that we change the sign of the term
$F^2/4\Lambda$ in eq. \SLa\ and that we replace $k$ with ${\rm sign}
(k)\,|k|.$ Radiative contributions to the local effective action
$\gm^{\rm loc}$ involving an odd number of epsilons then pick an
overall factor ${\rm sign}(k).$ This implies in particular that the
shift $\a$ for $k<0$ becomes ${\rm sign }(k)$ times the value of $\a$ for
$k>0.$

It is very easy to see that the type of shifts (iii) can not be
reproduced using a local higher covariant derivative term. Indeed,
some simple dimensional analysis shows that a local higher covariant
derivative parity-odd term never has the same mass dimension as a
local higher covariant derivative parity-even term.  Hence, both terms
can not be linearly combined into a single local term, which is the
requirement in (iii).

\smallskip

\noindent{\bf Acknowledgements:} FRR was supported by FOM, The
Netherlands. The authors also acknowledge partial support from
CICyT, Spain.

\smallskip

\refoutlw

\bye